\documentclass[useAMS,usenatbib]{mn2e}
\usepackage{graphicx,graphics,amsmath}
\newcommand{\vep}{\varepsilon}
\newcommand{\ep}{\epsilon}
\newcommand{\sig}{\sigma}
\newcommand{\Om}{\Omega_{\rm M}}

\newcommand{\LCDM}{{\Lambda}{\rm CDM}} 
\newcommand{\Ol}{\Omega_{\Lambda}}

\def\dchisq{\Delta_{\chi^2}} 
\def\hstkp{HST Key Project }
\def\degrees{\,^{\circ}}

\def\etal{{\em et al.}}

\def\apj{{Astroph.\@ J.\ }} 
\def\apjl{{Astroph.\@ J.\@ Lett.\ }} 
\def\aj{{Astron.\@ J.\ }}

\def\mn{{Mon.\@ Not.\@ Roy.\@ Ast.\@ Soc.\ }}

\def\na{{New \@ Astronomy \ }}

\def\pasp {{Pub.\@ Astro.\@ Soc. \@ Pacific }}

\begin{document}
\title {Non-Gaussianity and direction dependent systematics in HST key project data} 

\author[Shashikant Gupta and Tarun Deep Saini]
{Shashikant Gupta $^{1,2,3}$ and Tarun Deep Saini $^{2,4}$\\
  $^{1}$ Raman Research Institute, Bangalore, Karnataka, India, 560 080 \\
  $^{2}$ Indian Institute of Science, Bangalore, Karnataka, India, 560 012 \\
  $^{3}$ shashikant@physics.iisc.ernet.in \\
  $^{4}$ tarun@physics.iisc.ernet.in \\
} \maketitle

\begin{abstract}
Two new statistics, namely $\Delta_\chi^2$ and $\Delta_\chi$, based on extreme 
value theory, were derived in \cite{gupta08,gupta10}. We use these statistics 
to study direction dependence in the HST key project data which provides the 
most precise measurement of the Hubble constant. We also study the non-Gaussianity 
in this data set using these statistics. Our results for $\Delta_\chi^2$ show 
that the significance of direction dependent systematics is restricted to well 
below one $\sigma$ confidence limit, however, presence of non-Gaussian features 
is subtle. On the other hand $\Delta_\chi$ statistic, which is more sensitive to 
direction dependence, shows direction dependence systematics to be at slightly 
higher confidence level, and the presence of non-Gaussian features at a level 
similar to the $\Delta_\chi^2$ statistic.
\end{abstract}

\section{Introduction} 
Hubble's observations (1929) can be approximated as 
$v \propto d \, ,$
where $v=cz$ is the velocity (toward or away from us) of the galaxy being 
observed. Apart from a very few nearby galaxies, 
redshifts are positive indicating 
that the galaxies are receding away from us. The velocity of recession being 
proportional to its distance from us is explained by invoking the expansion of 
the Universe. $H_0$, the constant of proportionality, is called the Hubble 
constant and it measures the rate of expansion at the present epoch. The Hubble 
constant enters into various cosmological calculations and its importance can 
never be underestimated. It decides the value of critical density $\rho_c$, the 
amount of matter and energy required to make the geometry of the Universe flat. 
By comparing $\rho_c$ to the observed density one can decide the geometry of the 
Universe. Most importantly it sets the age of the Universe ($t_0$) and hence, 
size of the observable universe ($R_{ob}=ct_0$). Due to its importance 
determining its accurate value is of paramount importance.

Accurate measurement of the Hubble constant can also lead to test the cosmological 
principal (CP hereafter). According to CP the universe is homogeneous and 
isotropic at any given cosmic epoch.  If the cosmological principal is valid 
than one would expect the average value of Hubble constant to be same in 
different regions and in different directions. This issue has been addressed 
by various authors. We discuss some of the earlier results below. 
\\
{\bf{Are we living in a bubble? : }} 
If the matter distribution is not homogeneous, it causes variation in the value 
of Hubble constant. Since gravity pulls we expect that if a region of space has 
higher density then the average density the expansion rate will be negatively 
affected and thus the Hubble constant will be smaller in this region. In contrast 
to the local mass concentration a region with low mass density will produce larger 
value of $H_0$. \cite[]{zehavi98} provided the first evidence for a large local 
void. They measured $H_0$ within and outside $70\, h^{-1}$Mpc using SNe Ia to find 
that the value of $H_0$ was $6.5\, \%$ higher than that outside. This indicates a 
low density inner region compared to the outside one and is known as local bubble 
or Hubble bubble. The above authors had assumed a flat FRW universe with $\Om=1$ 
in their analysis. The variation in the inside and outside values of $H_0$ 
decreases to $4.5\, \% $ in the  $\LCDM$  model ($\Om =0.3$, $\Ol=0.7 $), however, 
it does not disappear completely. Recently \cite[]{jha07} revisited the problem using 
the latest SNe Ia data set and detected of the local Hubble bubble at $3 \sigma$ 
confidence level. However in a later publication \cite[]{conley07} claimed that  
it was a misinterpretation of color excess of supernovae. At this juncture it is 
difficult to say if the evidence for the local bubble is conclusive. 
\\
{\bf Variation in $H_0$, from HST key data :} \cite[]{mccl07} used HST Key Project 
data (see \S~\ref{sec:hstkey}) to calculate the variation in $H_0$ value. The authors 
find that a statistically significant variation in $H_0$ of 
$9 \, {\rm km \, s^{-1} \, Mpc^{-1}}$ exists in HST Key Project data. The approximate 
directional uncertainty is  $ 10\degrees $ to $ 20\degrees $. Their results indicate 
two sets of extrema that dominate on different distance scales. They find differences 
as great as $ \sim 35 \, {\rm km \, s^{-1} \,  Mpc^{-1} } $ within and 
$ \sim 20 \, {\rm km \, s^{-1} \, Mpc^{-1}} $ beyond our super-cluster. Within 
$ 70 \, {\rm Mpc} $ their results show a statistically significant difference of 
$ \sim 19 \, {\rm km \, s^{-1} \, Mpc^{-1} } $. This variation does not appear to be 
an artifact of Galactic dust, since there is no consistent difference looking in or out 
of the plane of the Galaxy. In fact, the overall structure in the map is inconsistent 
with the distribution of dust in the COBE dust maps \cite[]{schlegel98}. 

At this point one pertinent question is to ask ``are these variations in the 
measurement of $H_0$ due to a real departure from the cosmological principal?" 
On the contrary it is also possible that the data itself has some systematic 
errors due to some non corrected physical processes in the universe or there 
could be some real issues with the data reduction/calibration process. In order 
to comment on ``what is the real cause of the variations?", a critical review 
of the measurement methods is required. Measuring accurate value of the Hubble 
constant is a challenging task. One requires accurate measurement of redshift 
$z$ and the distance $d$. Although redshift can be measured with good accuracy 
from the spectrum of the light emitted by the object, the distance measurement 
is difficult. Various methods are employed to measure distances, namely 
Tully-Fisher relation (TFR), Surface brightness (SB) fluctuations, Fundamental 
plane (FP) relation, SNe type II, SNe type Ia, Sunyaev-Zeldovich effect (SZE), 
Gravitational lensing etc. Unfortunately most of these methods suffer from the 
systematic effects arising from many different causes. For instance SZE requires 
the 3-d distribution/shape of the plasma (hot gas) in the galaxy clusters. Radio 
and x-ray images of the clusters provide only the projected x-ray surface 
brightness and CMB decrement. Hence simplified assumptions about the shape of 
the cluster are made. Again the assumption about the phase and the temperature of 
the plasma are ad hoc \cite[]{sulk99}. In another example the uncertainty in the 
physical basis of Tully-Fisher relation can cause subtle systematic variances in 
the TF relation with environment. A critical review on the subject of $H_0$ 
measurement methods can be found in \cite{jacoby92}. 

Since there are many sources of systematics, special attention was needed to 
measure the accurate value of $H_0$. The most accurate experiment to achieve 
this was the Hubble Space Telescope (HST) key project \cite[]{hstkey}. We shall 
discuss the HST key project in the next Section, however the issues with this 
data have been mentioned in previous paragraphs. In the present paper we intend 
to put constraints on the CP using the $H_0$ data. This can be achieved by looking 
for direction dependent signatures in the data. As has been pointed out earlier 
detecting the direction dependent signatures does not guaranty departure from 
isotropy and hence CP. In that case one can constrain the reliability of the data. 
We use a technique \cite[]{gupta08,gupta10}, based on extreme value theory to 
accomplish this. Another important issue with any data set is the presence of 
non-Gaussian errors. Since Central Limit Theorem (CLT) predicts that the errors 
in the data should be Gaussian, the non-Gaussian errors are undesired and may 
indicate some unresolved issues. As a by product of our method we are able to 
detect the presence of non-Gaussian features in the errors, which makes our 
methods useful not only in the case of $H_0$ data but for any data set in general.

Plan of this paper is as follows. We discuss HST key project in \S~\ref{sec:hstkey}, 
which is the main source of our data. We discuss our methods in \S~\ref{sec:method}. 
Since our techniques are based on Extreme Value Theory (EVT hereafter) we discuss 
it briefly in \S~\ref{sec:evt}. Results and conclusions are presented in 
\S~\ref{sec:result} and \S~\ref{sec:conclusion} respectively. 

\section{HST Key Project and the Data set} 
\label{sec:hstkey} 
Measuring an accurate value of $H_0$ was one of the motivating reasons for 
building the NASA/ESA Hubble Space Telescope (HST). Measurement of $H_0$ with 
the goal of 10\% accuracy was designated as one of three ``Key Projects"  of 
the HST \cite[]{aaronson86,kennicutt95}. The overall goal of the HST Key Project 
was to measure H0 based on a Cepheid calibration of a number of independent, 
secondary distance determination methods. Many times the systematic errors 
dominate the accuracy of distance measurement. To overcome this the HST team 
averaged over the systematics and used a number of different methods to measure 
distances instead of relying on a single method alone. 

Determining $H_0$ accurately requires the measurement of distances far enough 
away so that both the small- and large-scale motions of galaxies become small 
compared to the overall Hubble expansion. To extend the distance scale beyond 
the range of the Cepheids, a number of methods that provide relative distances 
were chosen. The HST Cepheid distances were used to provide an absolute distance
scale for these otherwise independent methods, including the Type Ia supernovae, 
the Tully-Fisher relation, the fundamental plane for elliptical galaxies, surface 
brightness fluctuations, and Type II supernovae. The final result of HST key 
project \cite[]{hstkey} was $ H_0 = 72 \pm 8 \, {\rm km/s/Mpc} .$  
\\
{\bf{Data Set :}} 
We have chosen data from the HST key project \cite[]{hstkey} as our primary data 
set. This set contains 74 data points and provides a reasonably full sky coverage. 
Different methods used in order to this data set are : The TF relation, the FP 
relation, the SB fluctuations and SNe type Ia, SNe type II. In addition we have 
chosen 2 data points from \cite[]{sak00}. In all the cases recessional velocities 
have been corrected to the CMB frame and thus all the $H_0$ values belong to CMB frame. 
The full data set is published in table 1 of \cite[]{mccl07}. 

\section{Extreme value theory}
\label{sec:evt} 
In the present paper we investigate the direction dependent systematic effects, 
which exhibit anisotropy in the cosmological data. We identify the direction where 
the effect of the systematics is maximum. To estimate its statistical significance 
we need to know the distribution of this maximum, which can be computed using 
Extreme Value Theory (EVT). Since it is not a common tool in the arsenal of 
astronomers, we begin with a brief introduction of EVT. It was developed in 
parallel to the Central Limit Theory (CLT). While CLT describes limiting distribution 
of partial sums, EVT describes how the distribution of extremes (maxima/minima) 
looks like. Below we discuss the theory of maxima, however, the results obtained 
can be easily reformulated to obtain the distribution of minima. 

\begin{figure}
\centering
\includegraphics[angle=0,height=0.60\textwidth]{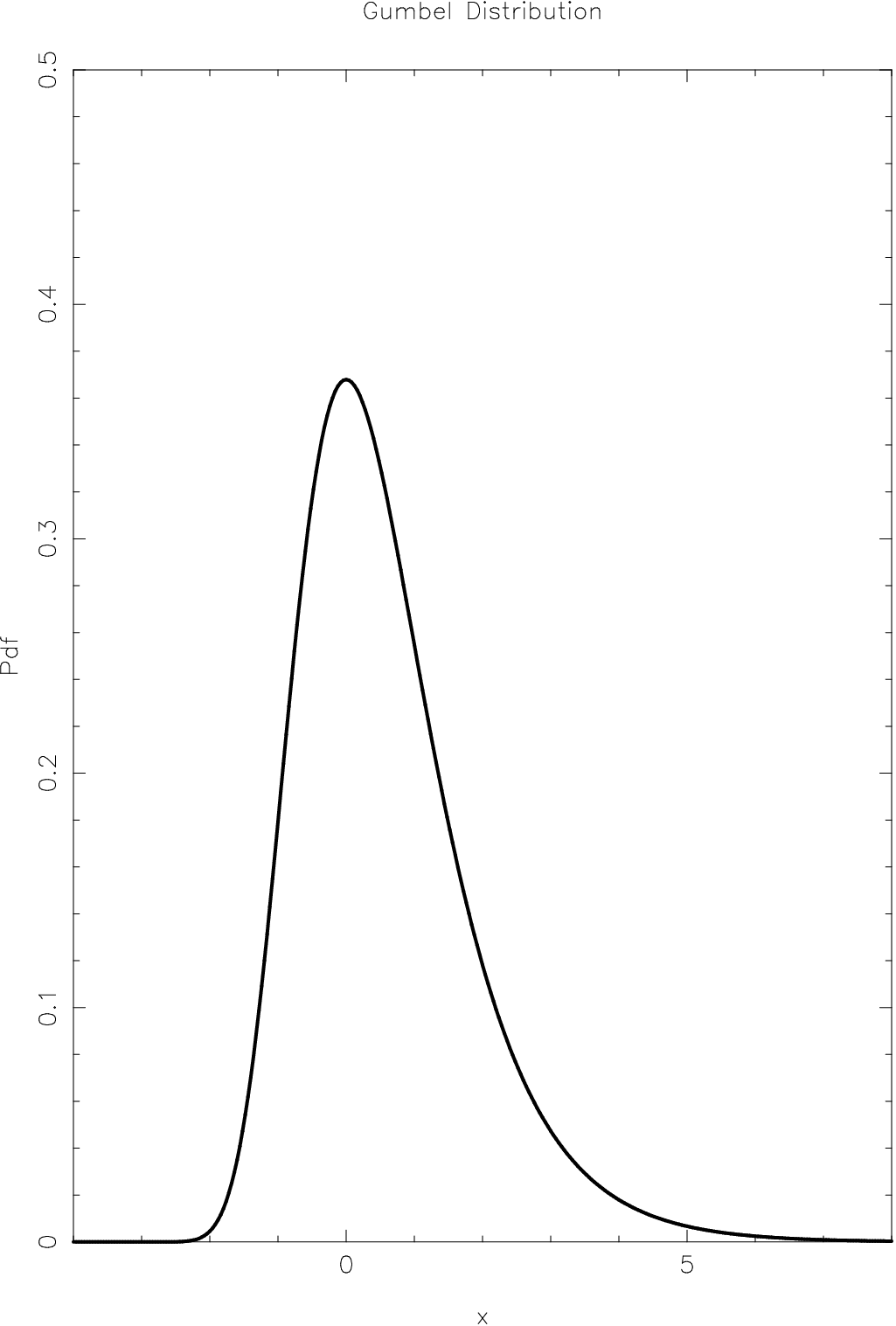}
\caption{A sketch of Gumbel distribution. Here we have plotted the probability 
distribution function which is derivative of $G(x)$. }
\label{fig_evt:gum}
\end{figure}

We shall outline the basic ingredients to obtain the theoretical distribution. 
Let F be a distribution with its right end point $x^{\star}$ which may be 
infinite, i.e. 
$$x^{\star} = {\rm sup}\{x:F(x)\le 1 \} \ . $$ 
We randomly choose a sample ($X_1, X_2, \cdots, X_n $) of size $n$ from this 
distribution. The maximum of this sample will approach $x^\star$ for large $n$ i.e. 
$ {\rm max}( X_1, X_2, \cdots ,X_n ) \rightarrow x^\star $ as $ n \rightarrow \infty $. 
The distribution of maxima has the following probability 
\begin{eqnarray}
&P({\rm max}( X_1, X_2, \cdots ,X_n ) \leq x)  \nonumber \\
& = P(X_1 \leq x, X_2 \leq x \cdots ,X_n \leq x)  \nonumber \\
& =  P(X_1 \leq x)P(X_2 \leq x) \cdots P(X_n \leq x) \nonumber \\ 
& =  F^n(x) \nonumber \  ;
\end{eqnarray}
where $X_1, X_2, \cdots , X_n$ are assumed to be independent. This converges to 
zero for $x < x^{\star}$ and to unity for $x \geq x ^\star$, which means that 
$F^n(x)$ is a degenerate distribution. A normalization of variable 
$ {\rm max}(X_1, X_2, \cdots ,X_n)$ is required in order to get a non-degenerate 
distribution. We choose linear normalization \cite[]{haan}. Let us assume that 
there exists a sequence $a_n>0$, and $b_n$ real such that 
$$ t= \frac{ max(X_1, X_2, \cdots ,X_n) - b_n}{a_n} $$ 
has a non-degenerate limiting distribution as $n \rightarrow \infty $, i.e. 
\begin{equation} 
\lim_{n \rightarrow \infty} F^n(a_nx + b_n) = G(x) \ , 
\end{equation}
where $0 \le G(x) \le 1 $, $G(x)$ is the required distribution. It is very 
difficult to derive the general form of $G(x)$, we mention some of the results 
available in literature. We state a theorem due to Fisher and Tippett \cite[]{fish28} 
which describes the required distribution. 
\newtheorem{fisher}{Theorem}
\begin{fisher}
The distribution of maxima $G(x:\vep)$ has the following form :
\begin{equation}
G(x; \vep ) = \exp \left( -\left[ 1 + \vep x  \right]^{-1/\vep}  \right) \ , 
\label{eq:gx} 
\end{equation}
with $1+\vep x > 0$. Where $\vep \ep R$ is called the shape parameter or the 
extreme value index. 
\end{fisher} 
Proof of this theorem \cite[]{fish28} is beyond the scope of this paper. 
We only outline a few interesting facts :
\begin{itemize} 
\item When $\vep=0$ : \\
Tailor expansion of 
$ \left[ 1 + \vep x  \right]^{-1/\vep}$ gives $e^{-x}$. Thus Eq~\ref{eq:gx} gives 
$$G(x) = \exp(-e^{-x}) \, . $$ 
This is known as the Gumbel distribution or extreme value distribution of type I. 
The right end point of this distribution is infinity. Also $1-G(x) \sim e^{-x} $ 
as $x \rightarrow \infty $, indicating that the distribution has a thin tail. It 
is clear from the form of Gumbel that it is unbounded on either side that is, there 
is no $x$ for which $G(x)=0$. A sketch of Gumbel distribution is shown in 
Fig~\ref{fig_evt:gum}. 

\item When $\vep>0$ : \\
$ G(x;\vep) < 1$ for all $x$, i.e. the right end point of the distribution is 
infinity. Also as $x \rightarrow \infty $ , $1-G(x,\vep) \sim  (\vep x)^{-1/\vep} $. 
Which indicates that the distribution has a heavy tail. This is called Frechet 
distribution, or type II extreme value distribution. One can clearly see that 
this distribution has a lower bound. 
\item When $\vep<0$ : \\
The right end point of the distribution is $-1/\vep$. This is known as Weibull 
distribution, or extreme value distribution of type III. 
\end{itemize}

In the above discussion we have not mentioned the location and scale parameters 
for simplicity. When we introduce these parameters, the form of $G(x)$ becomes 
complicated as shown below 
\begin{equation}
G(x;m ,s, \vep ) = \exp \left( -\left[ 1 + \
\vep (\frac{x-m}{s})   \right]^{-1/\vep}  \right) \ , 
\end{equation}
where $m =b_n \in R $ is the location parameter and $s=a_n > 0$ is the scale parameter. 

For our analysis we shall use the Gumbel distribution since the variable of interest 
there does not have bound on either side. Thus our emphasis will be on the  Gumbel 
distribution in the rest of this paper, which with scale and location parameters 
takes the form 
\begin{equation}
\label{eq:gumbel}
G(x;m ,s) = \exp \left( -\left[ \exp - (\frac{x-m}{s}) \right]   \right)  \ , 
\end{equation} 
The probability distribution (pdf) can be derived by differentiating Eq~\ref{eq:gumbel}. 

\section{Methodology}
\label{sec:method} 

Two new techniques, namely $\dchisq$ and $\Delta_\chi$ statistics based on Extreme 
value theory, were derived in \cite{gupta08,gupta10}. We apply the same method 
here, however we briefly mention these techniques for completeness. 

\subsection{$\Delta_{\chi^2}$ statistic}
\label{sec:method_chisq} 
First we calculate the best fit value  of $H_0$ for the \emph{complete data set}  
by minimizing $\chi^2$, defined below 
\begin{equation}
\chi^2 = \sum_i \left( {\frac{H_{0i} - H_{0} }{\sig_i}}\right) ^2 \, ,
\label{eq:H0_bf} 
\end{equation}
where $ H_{0i} $ is the $i^{th}$ point in the data set and $\sigma_i$ is the 
observed standard error. This minimization gives us the best fit value $H_{0b}$. 

Using this best fit value we now define $\chi_i$ for the $i^{th}$ data point as 
\begin{equation}
\chi _i = {\frac{H_{0i} - H_{0b} }{\sig_i}} \ , 
\label{eq:chi} 
\end{equation}
where $H_{0b}$ is the best fit value of the Hubble constant. We shall consider 
subsets of the data set to construct our statistic. We define the reduced $\chi^2$ 
in terms of $\chi_i$ as follows 
\begin{equation}
\chi^2= \frac{1}{N_{\rm subset}}\sum_i \chi_i^2 \,\,,
\label{eq:chisq}
\end{equation}
where it should be noted that by `reduced' we do not mean `per degree of freedom', 
since we \emph{do not fit} the model separately to the subsets of the data. 
Here $\chi^2$ is an indicator of the statistical scatter of the subset from the 
best fit value $H_{0b}$.

If the Cosmological Principle holds then the value of the Hubble constant should 
not depend upon the direction in which it is measured. We use this fact to choose 
specific subsets of data. We divide the complete data into two hemispheres, labeled 
by the direction vector $\hat{n_i}$, and take the difference of the reduced 
$\chi^2$ computed for the two hemispheres separately to obtain 
$\Delta \chi_{\hat{n_i}}^2 =\chi^2_{\rm north} - \chi^2_{\rm south} $, where we 
have defined 'north' as that hemisphere towards which the direction vector 
$\hat{n_i}$ points. We are only interested in the magnitude of this difference, 
therefore, we take the absolute value of $\Delta \chi_{\hat{n_i}}^2$, and then vary 
the direction $\hat{n}$ across the sky to obtain the maximum absolute difference 
\begin{equation}
 \dchisq = {\rm max} \{| \Delta \chi_{\hat{n_i}}^2 |\}\,\,. 
\end{equation}
We note that since the same data point appears in several subsets, the maximization 
is not done over statistically independent measures of our statistic. Another noteworthy 
fact is that if the direction dependence has a forward-backward symmetry then this 
statistic will not be able to detect it. However, due to its ease of construction and 
use we consider this simplest of possible statistics.

To interpret our results we need to know the range of $\dchisq$ that we can expect 
if there were no direction dependence in data, and the noise in the measurements 
were Gaussian. The spatial distribution of measurements is not uniform on the sky, 
therefore, the number of measurements in the two hemispheres, for a given direction, 
varies with the direction $\hat{n}$ in a complicated manner. Therefore one might 
expect the probability distribution function $P(\dchisq)$ to be extremely complicated, 
however, extreme value theory, \S~\ref{sec:evt}, shows that the distribution is, 
in fact, a simple, two parameter Gumbel distribution, characteristic of extreme 
value distribution type~I: 
\begin{equation}
\label{eq:delta_gumbel}
P(\dchisq) =\frac{1}{s} \exp \left[ -\frac{\dchisq-m}{s}\right]\,\exp\left
  [-\exp\left(-\frac{\dchisq-m}{s}\right)\right]\,\,, 
\end{equation} 
where the position parameter $m$ and the scale parameter $s$ completely determine 
the distribution.

To quantify departures from isotropy we need to know the theoretical distribution 
$P_{\rm theory}(\dchisq)$. Even though we know what to expect in a general manner, 
it is difficult to obtain the parameters $s$ and $m$ analytically, therefore, we 
calculate this distribution numerically by simulating several sets of Gaussian 
distributed $\chi_i$ on the measurement positions and obtaining $\dchisq$ from each 
realization. 

If the noise in the data is Gaussian then the above distribution adequately 
quantifies the directional dependence in the data. But if the data has 
non-Gaussian noise then the theoretical distribution cannot be used to 
quantify the level of significance of our possible discovery of anisotropy. 
We construct an independent test for directional dependence by obtaining 
the bootstrap distribution $P_{\rm BS}(\dchisq)$, which is constructed in the 
following manner.  The observed $\chi_i$'s are assumed to be drawn from 
some unknown, direction dependent probability distribution. We shuffle the 
data values $H_{0_i}$ and $\sigma_i$ over the measurement positions, 
thus destroying any directional alignment they might have had due to anisotropy. 
Thus we are able to generate several realizations of data and estimate the 
distribution $P_{\rm BS}(\dchisq)$.

\begin{figure}
\centering
\includegraphics[ width=0.5\textwidth]{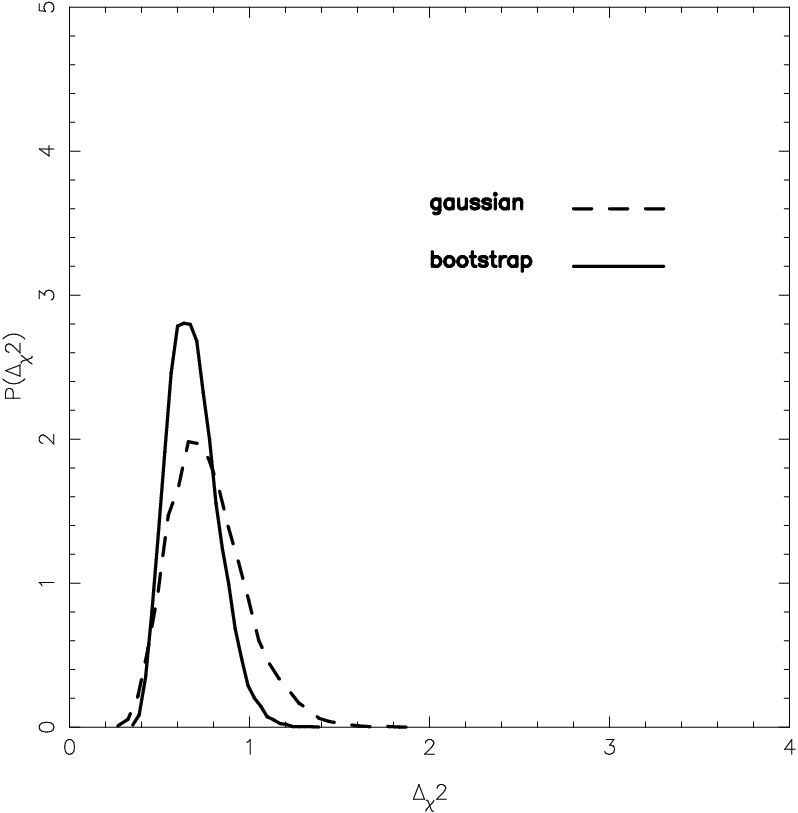} 
\caption{ 
Here we plot the probability density of $\dchisq$ for the simulated data. Solid 
curve represents the bootstrap distribution while broken curve represents the 
theoretical distribution assuming Gaussian errors.} 
\label{fig:mock}
\end{figure}

In order to know what to expect from this statistic we simulate data and 
calculate the above mentioned Bootstrap and numerical distributions. These 
are shown in Fig.~\ref{fig:mock}. As had been discussed in \cite[]{gupta10} 
there exists a specific bias between the two distributions. Since the numerical 
distribution is obtained by assuming $\chi_i$s to be Gaussian random variates 
with a zero mean and unit variance, therefore, it does not have any bounds. However 
the bootstrap distribution is obtained by shuffling through a 
\emph{specific realization} of $\chi_i$ where the $\chi_i$s are obviously bounded. 
It is clear that on the average this should produce slightly smaller values of 
$\Delta_{\chi^2}$ in comparison to what one expects from a Gaussian distributed 
$\chi_i$s. 

Our results for $\Delta_{\chi^2}$ statistic in this paper should be interpreted 
with respect to Figure~\ref{fig:mock}. Concerns regarding the small number of data 
points in data set and its effect on the efficacy of our method can be addressed 
(as in \cite{gupta10}) by noting that this figure is produced with only $76$ points 
and the theoretical and the bootstrap distributions look similar.

\subsection{ $\Delta_{\chi}$ statistic}
\label{sec:method_chi}
As mentioned above, $\chi_i^2$ does not contain information about whether the 
measurement is above or below the fit i.e. greater or smaller than $H_{0b}$.  
An obvious generalization that does contain this information can be obtained 
by considering a statistic based on $\chi_i$s. We consider two subsets of data 
defined by two hemispheres labeled by the direction vector $\hat{n}$, 
containing $N_{\rm north}$ and $N_{\rm south}$ data points, where the total number 
of data points,  $N = N_{\rm north} + N_{\rm south}$,  and define the quantity 
\begin{equation}
\Delta\chi_{\hat{n}} = \frac{1}{\sqrt{N}} \left( \sum_{i=1}^{N_{\rm north}} \frac{\chi_i}{\sigma_i} - \sum_{j=1}^{N_{\rm south}} \frac{\chi_j}{\sigma_j} \right )\,.
\end{equation}
Clearly $\langle \Delta \chi_{\hat{n}} \rangle = 0$ and $\langle 
(\Delta \chi_{\hat{n}})^2 \rangle = 1$. From the central limit theorem \citep{ken77} 
it follows that for $N \gg 1$, the quantity $\Delta\chi$ follows a Gaussian 
distribution with a zero mean and unit variance. As in the previous case we maximize 
this quantity by varying the direction $\hat{n}$ across the sky to obtain the maximum 
absolute difference 
\begin{equation}
\Delta_{\chi} = {\rm max} \{| \Delta \chi_{\hat{n}} |\}\,\,. 
\end{equation}

This statistic differs from the previous one in that the $\Delta_{\chi}$ statistic 
has a theoretical limit where the position and the shape parameters can be determined 
analytically. Given $N_d$ independent directions on the sky we are essentially 
determining the maximum of a sample of size $N_d$ where the individual numbers are 
drawn from a Gaussian distribution with a zero mean and unit variance. In the limit 
$N_d \gg 1$ the parameters are given by \citep{haan} 
\begin{eqnarray}
m &=& \sqrt{2\log N_d -\log \log N_d - \log 4\pi}\\
s &=& \frac{1}{m}
\end{eqnarray}
where we have to additionally assume that the number of measurements $N\gg 1$, since 
the distribution for $\chi$ becomes Gaussian only in this limit. This is convenient 
since at least for large data sets, which will be available in the future, a comparison 
with theory becomes simpler. However, for a smaller number of measurements (data points) 
there is a possibility that not all directions are independent, in fact, it is quite 
possible that two directions contain exactly same subsets in the two hemisphere. In 
this situation is is clear that the total independent directions is a smaller number 
than $N_d$ and thus theoretical distribution would be rightward shifted and also more 
sharply peaked. For this reason we also calculate the bootstrap distribution and the 
theoretical distribution in the same manner as for the previous statistic.

\section{Results}
\label{sec:result}

\begin{table}
\begin{center}
\caption{Best fit value for $H_0$ 
 \label{tbl:h0_bf}} \bigskip 

\begin{tabular}{cccc}
\hline
Best fit  & $\chi^2$ & $\chi^2_{\rm per \, dof} $ \\
\hline
72.0  & 194.1 & 2.6 \\
\hline\hline
\end{tabular}
\end{center}
\end{table}

\begin{figure}
\centering
\includegraphics[ width=0.5\textwidth]{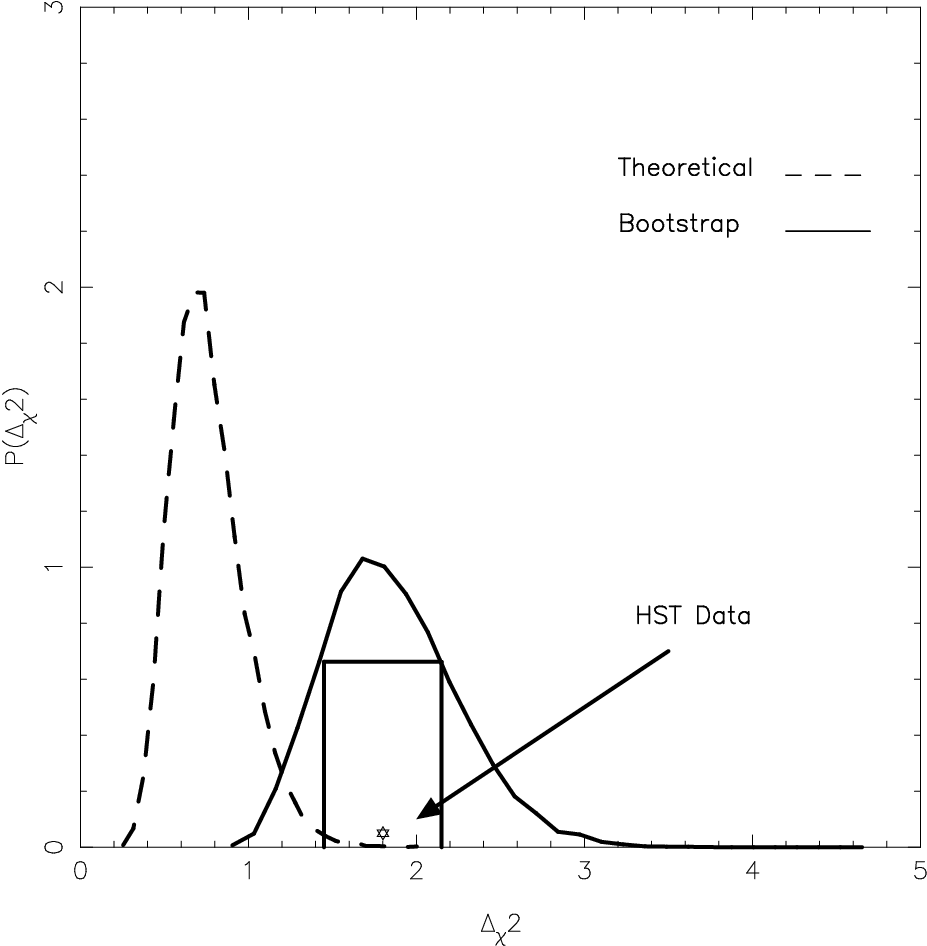} 
\caption{Here we plot 
the probability density of $\dchisq$. Solid curve represents the bootstrap 
distribution while broken curve represents the theoretical distribution assuming 
Gaussian errors. The two curves do not match. $\dchisq$ for the data lies close 
to the peak of bootstrap distribution.} 
\label{fig:pdf_chisq}
\end{figure}

First we obtain the best fit value, $H_{0b}$, of the Hubble constant for the data, 
which is shown in Table~\ref{tbl:h0_bf}. The large value of $\chi^2$ suggests 
that the error bars may have been underestimated. 

{\bf{$\dchisq$ statistic :}}
We have applied the $\dchisq$ statistic to this data and calculated the bootstrap 
and theoretical distributions for the HST Key Project data, as discussed earlier. 
The theoretical and bootstrap distributions for this data are shown in 
Fig~\ref{fig:pdf_chisq}. To interpret our results we compare with Fig~\ref{fig:mock}. 
We see that the Theoretical distribution has been shifted far away on the left side 
of the bootstrap distribution. This contradicts to the fact that bootstrap 
distribution should lie slightly to the left of the theoretical distribution, as 
explained above, indicating the non-Gaussian nature of errors in the data. 
HST Key data lies outside the theoretical distribution, but is close to the peak of 
the bootstrap distribution. This suggests that either the data is free from direction 
dependent systematics or a more sensitive technique is required to put constraints on 
these systematics. 

{\bf{$\Delta_\chi$ statistic :}} 

\begin{figure}
\centering
\includegraphics[ width=0.5\textwidth]{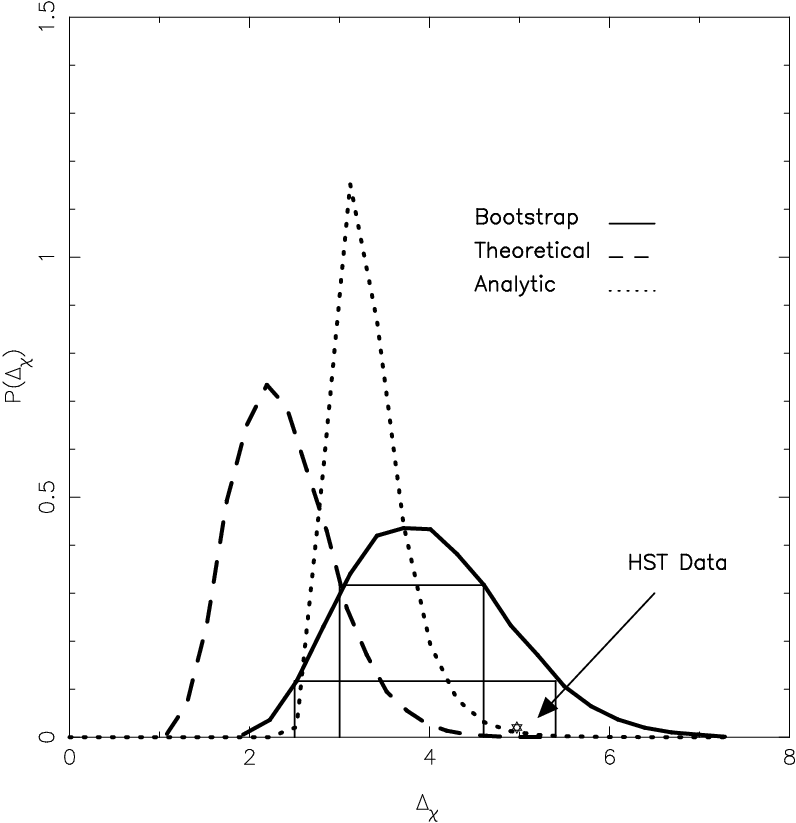} 
\caption{Here we plot the probability density of $\Delta_\chi$. Solid curve represents 
the bootstrap distribution while broken curve represents the numerically calculated 
distribution assuming Gaussian errors. The two curves do not match very well. The 
analytically calculated distribution is shown by dotted lines. $\Delta_\chi$ for the 
HST key data lies outside one $\sigma$ region around the peak of bootstrap distribution.} 
\label{fig:pdf_chi}
\end{figure}

We apply $\Delta_\chi$ statistic to the data and calculate the bootstrap and numerical 
distributions. These are shown in Fig~\ref{fig:pdf_chi}. Here also the numerical 
distribution lies to the left of the bootstrap distribution, which is in contradiction 
to the expected relative positions of the two. Interestingly, in this case we have an 
advantage, since we can calculate the theoretical distribution analytically. This 
analytic distribution is shown by dotted line in Fig~\ref{fig:pdf_chi}. It was discussed 
in \S~\ref{sec:method_chi} that if all the directions are not independent then the 
analytic distribution should lie on the right side of the numerical distribution and 
should be peaked sharply. This is what we observe in Fig~\ref{fig:pdf_chi}. 
Bootstrap distribution lies to the right of the analytic distribution, however, 
we find that it is wider than what we would expect from a true Gumbel distribution. 
This indicates that the bootstrap distribution is not truly Gumbel, therefore, it 
should be compared with the numerical distribution, which is calculated in a manner 
that is identical to the bootstrap one. Comparison shows that the two distributions 
differ in a manner identical to the difference seen in Fig~\ref{fig:pdf_chisq}, 
based on the $\Delta_{\chi^2}$ statistic; indicating a similar level of non-Gaussianity. 
We also find that the position of the data lies outside the 1 $\sigma$ region from the 
peak of the bootstrap distribution. 

\section{Conclusions}
\label{sec:conclusion}
We have applied the $\dchisq$ and $\Delta_\chi$ statistics to the \hstkp data. We 
find that in both the cases the bootstrap and the theoretical distributions are 
very different from each other. Thus we conclude that the nature of the errors in 
the data is non-Gaussian. $\dchisq$ statistic does not show direction dependence in 
the data, however, $\Delta_\chi$ statistic which is more sensitive to the direction 
dependence, shows the presence of direction dependent systematics at around one 
$\sigma$ level. 

{\bf{Acknowledgements}} Shashikant thanks Arnab Rai Chodhuri for supporting part of 
this work by his DST grant (DST0815). 


\end{document}